# Replay-Augmented Regularisation for Class-Incremental Network Intrusion Detection: A Three-Dataset Evaluation of EWC and LwF

**Mohammed Fackruddin Tusher[1],***
Department of CSE, BRAC University, Dhaka, Bangladesh
Email: fackruddin.tusher@gmail.com
**Muhammad Iqbal Hossain, PhD[2]**
Associate Professor, Department of CSE, BRAC University, Dhaka, Bangladesh
Email: iqbal.hossain@bracu.ac.bd

**Abstract:** As new attack categories emerge, network intrusion detection models need to be updated; however, when fine-tuned sequentially on new classes, previous classes are forgotten, resulting in catastrophic forgetting. The three public datasets we use are UNSW-NB15, ToN-IoT and CIC-IDS2017, and for these datasets, our class-incremental intrusion detection system is built using a CNN with one-dimensional (1D) archi- tecture trained over nine sequential tasks, each of which introduces one attack category. We evaluate sequential fine-tuning without any buffer compared to Elastic Weight Consolidation (EWC) and Learning without Forgetting (LwF), both of which include a class-size-adaptive rehearsal buffer of 20% of all previous classes (from 20 to 300 samples). The hyper-parameters were selected using UNSW-NB15 and fixed for the other two datasets. On UNSW-NB15, replay improves the final accuracy of both regularisers to 73.6% (EWC) and 71.1% (LwF) compared to about 32% (no improvement over fine-tuning). LwF with replay achieves 46.1% and 91.6% final accuracy for the two frozen-configuration datasets (ToN-IoT, CIC-IDS2017) while EWC with replay achieves 23.7% and 61.9%. In order to avoid this, EWC was therefore tuned on one dataset and the other two data sets were used as test sets. It is also demonstrated that the common forgetting measure can be misleading for a model which does not learn the old classes. As the fine-tuning baseline was not re- played and all the results were based on a single seed, it is not claimed that the regularisers provide any extra support over replay alone.

## 1 Introduction

Network intrusion detection system (NIDS): It is a system that monitors network traffic and alerts to potential intrusion. Amongst the broader defence, it is one of the layers, which also involves access control and incident response [1]. The attack surface is constantly evolving and new attack types are constantly being discovered and added to a deployed classifier. Retraining the network begins anew from all historical data can be costly, and fine-tuning only on the new data can cause catastrophic forgetting [2, 3] to the network, which overwrites previous knowledge of earlier classes. The behaviour of the model in Normal traffic is also in question as false alarms on normal traffic are very expensive in the operation [18]. Continual learning aims to find out how to learn from a stream of tasks without such a loss. There are typically 3 scenarios: task, domain and class-incremental learning [5]. The most difficult of the three is class-incremental where new attack types are added to a detector without it being told what attack type a particular flow is. There are two large families of methods in use: These regularisation-based approaches do not require any storage and either regulate the weights or the outputs, Elastic Weight Consolidation (EWC) [6] and Learning without For- getting (LwF) [7]. The replay based methods [9–11] store a small memory of previous samples and add it to the training. There are a number of continual learning benchmarks, primarily focused on image classification tasks [4, 5]. Both questions are not answered in practice for intrusion detection. Second, the extent to which replay budgets are scaled up with respect to the length of the task sequence.Third, the value added contribution of a small, bounded replay budget to regularisation based methods in a long task sequence with extreme class imbalance. Second,whether a configuration selected from one data set can be trusted when applied to another data set, as flow-level data sets vary in feature sets, capture environments, and class structure. Typical continual learning metrics can also be misleading if the model does not contain the old classes.

This paper addresses three research questions.

**RQ1** How much does a small class-adaptive rehearsal buffer improve final accuracy and retention for EWC and LwF?

**RQ2** Does a configuration selected on UNSW- NB15 transfer, without re-tuning, to ToN- IoT and CIC-IDS2017?

**RQ3** Which standard continual-learning metrics are informative in this setting, and which are misleading?

Our contributions are as follows.

- A reproducible nine-task class incremental protocol for flow-based intrusion detection, applied to three public datasets with one shared pipeline.

- A simple class-size-adaptive replay rule with a closed-form per-class budget (Eq. 6).

- A controlled ablation on UNSW-NB15 showing that EWC and LwF without replay are indistinguishable from fine-tuning, while the same methods with replay are not.

- A frozen-configuration evaluation on ToN-IoT and CIC-IDS2017, with an

analysis of where EWC fails and why forgetting and backward transfer can mislead.

## 2 Background and Related Work

**Catastrophic forgetting and scenarios.** Contextual, sequentially trained networks forget earlier information when later information doesn't contain it [2]. According to [3, 4] surveys categorize remedies into regularisation, replay and architectural approaches. The class-incremental setting requires separating all classes observed up to that point with a single output layer and without a task label, thereby making the bias towards current classes a major challenge [5].

**Regularisation.** EWC imposes a quadratic penalty on parameters that are deemed to be important from the Fisher information [6] and were influential in previous tasks. LwF records the soft predictions of the old model on new data and includes a distillation term [7, 8] to ensure that the new model continues to act like the old model when the outputs are the same.

**Replay.** A few repetitions of previous samples are good baseline, and rehearsing a few repetitions is essential to class-increase forgetting, but not sufficient [13].

**Metrics.** Standard, and many other proposed metrics, are accuracy over the learned tasks, forgetting [12] and backward transfer [10] are standard, and others have been proposed as additional metrics, because forgetting is not enough [13].

**Intrusion-detection-data.** Intrusion-detection data is generally employed with several types of attacks, such as the UNSW-NB15 dataset (2015) [14] and the ToN-IoT dataset (2022) [15] and the CIC-IDS2017 dataset (2018) [16]. They have reported experiences with quality problems: CIC-IDS2017 has experienced labelling issues (as well as flow-construction issues) documented in [17] and machine-learning evaluations of intrusion detection can easily be interpreted outside the closed world of a single dataset, as noted in [18]. The reason for that is that we test with three sets of data using the same
pipeline.

## 3 Methodology

Figure 1 explains how the three datasets pass through the pipeline.

### 3.1 Problem formulation

Let $K$= 10 be the number of classes (Normal plus nine attack categories) and T= 9 the number of tasks. An attack order (a1,...,aT ) is fixed for each data set. The sum of the class sets up to task t is

$$\mathcal{C}_t = \{\text{Normal}, a_1, \dots, a_t\} \qquad (1)$$

Training data for task t contains only the newly introduced attack class and Normal traffic,

$$\mathcal{D}_t^{\text{new}} = \{(x, y) \in \mathcal{D}_{\text{train}} : y \in \{a_t, \text{Normal}\}\} \qquad (2)$$

For that reason, normal traffic is fully available at all tasks, assuming that benign flows are plentiful, while old attack classes are not available at all tasks, aside from the rehearsal buffer. No task identity or logit masking is used at test time, and the model $f\theta : R^d \to R^K$ has a single output layer with all K logits at the beginning, thus the setting is class-incremental. A cumulative test set (evaluation set) for task j.

$$E_j = \{(x, y) \in D_{test}: y \in C_j\} \tag{3}$$

### 3.2 Model

The classifier is a one-dimensional residual CNN having approximately 0.23 M parameters. The input feature vector is treated as a one-channel sequence and passes through Conv1d(1 → 64, k=3)–BatchNorm–ReLU, Conv1d(64 → 128, *k=3*)–BatchNorm–ReLU, dropout (0.20), two residual blocks (two Conv1d(128 → 128, *k=3*) layers with BatchNorm, ReLU and dropout 0.15, plus an identity shortcut) [19, 20], global average pooling, a fully connected layer (128→64, ReLU, dropout 0.20) and a linear classifier (64 → 10). All the methods and datasets are initialized exactly the same and all follow the same architecture.

### 3.3 Class-weighted training objective

Within a task the training set is strongly imbalanced (for example, about 36 000 Normal flows against about 90 Worms flows in the last UNSW-NB15 task). Without correction all methods collapsed to predicting Normal (Section 5.2). We use a class-weighted cross-entropy. For the set of classes P present in the task's training data, with $n_c$ samples of class c,

$$w_c = \frac{1/n_c}{\frac{1}{|\mathcal{P}|}\sum_{c'\in\mathcal{P}} 1/n_{c'}}, \quad c \in \mathcal{P}; \qquad w_c = 0 \quad \text{otherwise} \tag{4}$$

The batch loss is

$$\mathcal{L}_{\mathrm{CE}}^{w} = -\frac{\sum_i w_{y_i} \log p_\theta(y_i \mid x_i)}{\sum_i w_{y_i}} \tag{5}$$

### 3.4 Class-size-adaptive rehears buffer

For each class c introduced before the task *t,* let $N_c$ be the number of its training samples. The replay budget for that class is

$$n_c = \min\{\max(\lfloor \gamma N_c \rfloor, n_{\min}), n_{\max}, N_c\}, \quad \gamma = 0.20, \quad n_{\min} = 20, \quad n_{\max} = 300 \tag{6}$$

The fraction γ scales the budget according to the size of the class, the floor $n_{min}$ ensures that rare classes are kept represented in the buffer, while the cap $n_{max}$ prevents overwhelming the buffer and the training time by large classes. A subset $R_c \subset D_c$ of $n_c$ samples is randomly sampled without replacement from the set of training samples $D_c$ of class c. The buffer at task t is the collection of classes that is the union of the previous classes,

$$\mathcal{B}_t = \bigcup_{c\in\mathcal{C}_{t-1}} \mathcal{R}_c, \quad \mathcal{B}_1 = \emptyset \tag{7}$$

and EWC and LwF are trained on $D^{new}_t \cup B_t$ . At most times, less than 8% of the last task samples in each dataset are stored in the buffer: $|C_{t-1}| \leq 2700$ for most of the time, which is 2700 samples < 300. The classes that are replayed include normal. The buffer is re-drawn at every task (with a fixed seed) and uses the training samples previously saved from the previous classes in our implementation and it is more of a per-task replay budget than strictly persistent memory (see section 7).

### 3.5 Elastic Weight Consolidation

After finishing task t, an empirical Fisher estimate is computed on The mini-batch gradients of the are squared and averaged to get Dnew $t$ . (unweighted) cross-entropy,

$$F_t = \frac{1}{B_t} \sum_{b=1}^{B_t} \left( \nabla_\theta L_{CE}^{(b)}(\theta_t^*) \right)^{\odot 2}, \tilde{F}_t = \sum_{s=1}^{t} F_s \quad (8)$$

where θ∗t are the parameters selected at the end of task t. Fisher estimates are totals of the tasks and the anchor which is the most recent parameter vector. Minimizes training in task t + 1

$$L_{EWC} = L_{CE}^{w} + \frac{\lambda}{2}\sum_{i} \tilde{F}_{t,i}(\theta_i - \theta_{t,i}^{*})^2, \lambda = 3 \quad (9)$$

and statistics of BatchNorm are not covered by the Penalty.

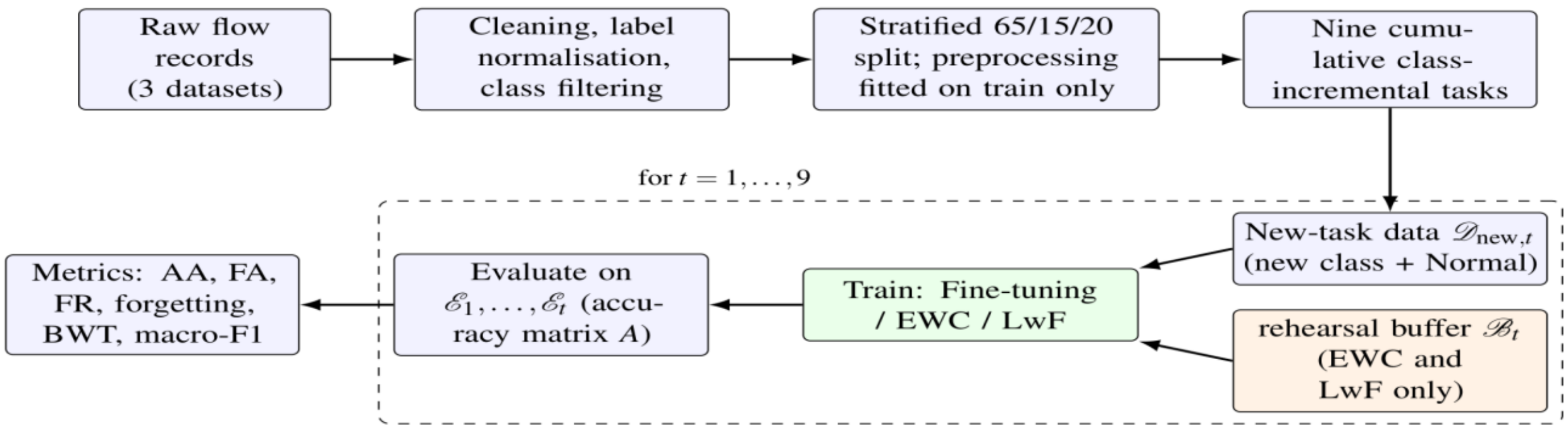

**Figure 1:** Experimental pipeline. All datasets and methods share data preparation. The model is trained for every task t using the data of the new-class. ( for EWC and LwF, a rehearsal buffer), and then tested on the All Test Sets for tasks seen so far.

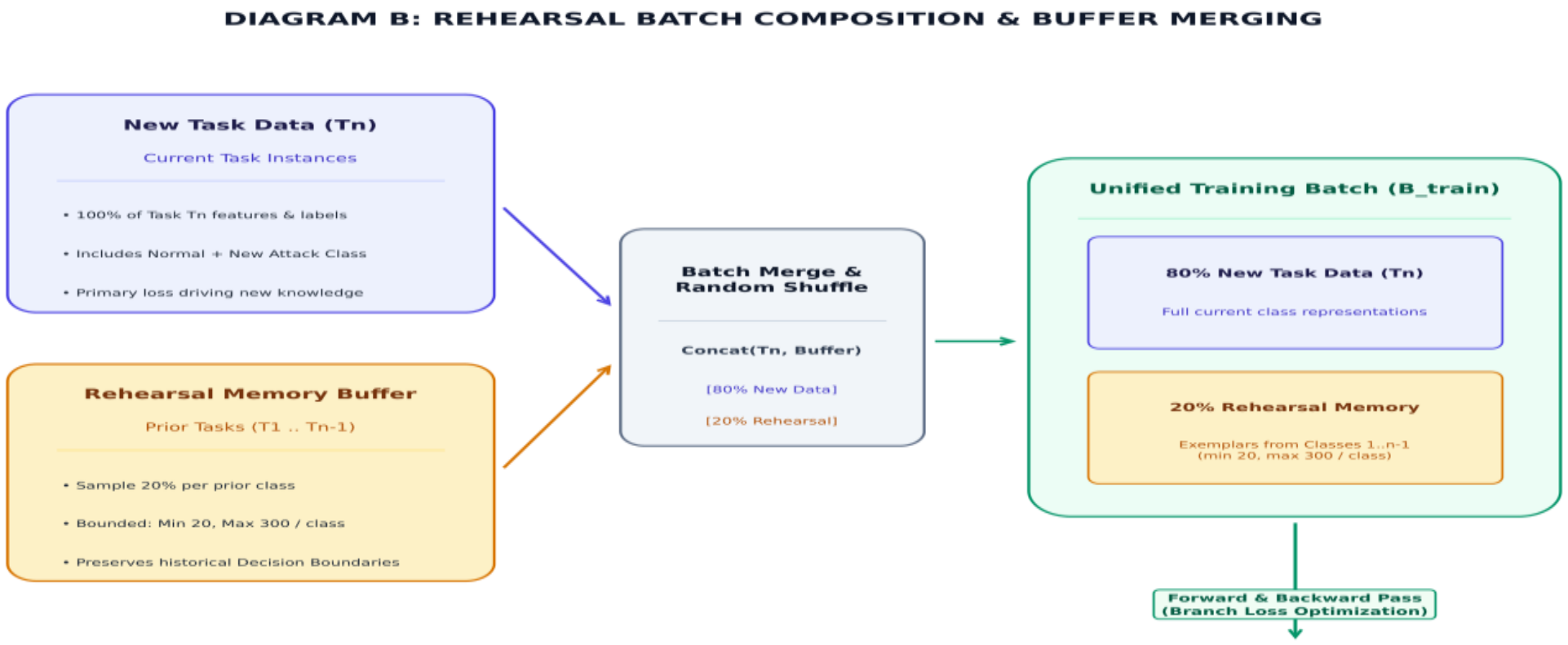

**Figure 2: Rehearsal Buffer**

### 3.6 Learning without Forgetting

Before task *t* a frozen copy of teacher is stored in the previous model. Suppose the teacher logits and teacher student logits are zT and zS, respectively, and are limited to the classes Ct−1, and τ the temperature.
The distillation loss and the total loss are

$$\mathcal{L}_{\mathrm{KD}} = \tau^2 \mathrm{KL}\left(\mathrm{softmax}(z^T/\tau) \,\|\, \mathrm{softmax}(z^S/\tau)\right), \quad \mathcal{L}_{\mathrm{LwF}} = \alpha \mathcal{L}_{\mathrm{CE}}^{w} + (1-\alpha)\mathcal{L}_{\mathrm{KD}}. \quad (10)$$

with $\alpha = 0.4$ and $\tau = 3$, applied to every sample in the batch, including replayed one

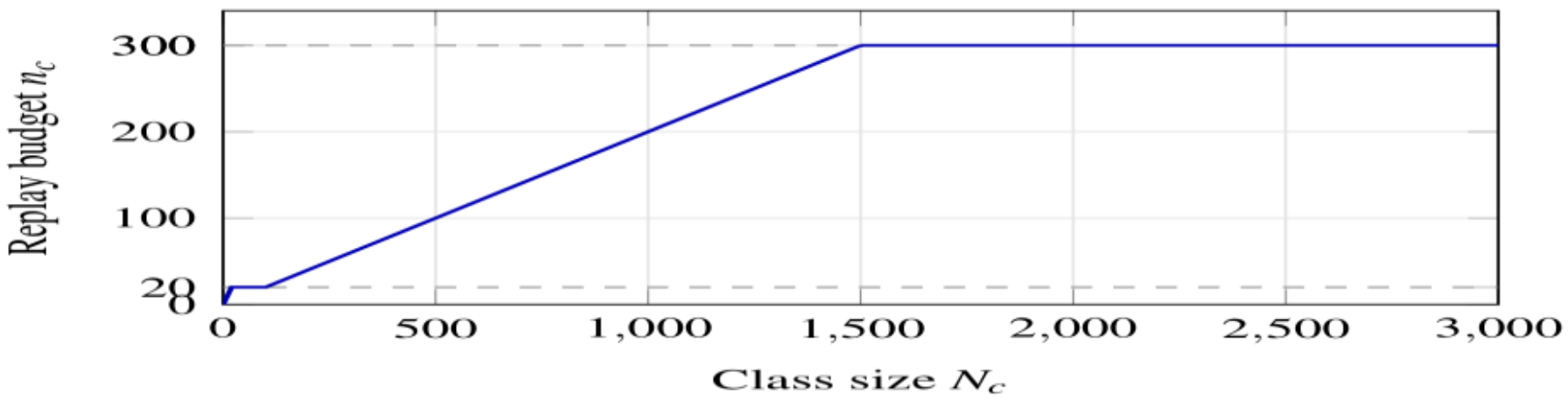

**Figure 3:** Replay budget per class (Eq. 6). Entirely kept small sized classes, mid-sized classes contribute 20%,and classes with more than 1 500 samples are capped 300.

### 3.7 Training procedure

Each task uses a fresh Adam optimiser [21] (learning rate 10−3, batch size 32) for at most 60 epochs. Early stopping with patience 10 monitors accuracy on the validation samples of the new class and Normal (it does not include old classes), and the best checkpoint is restored. All methods start from the same seed, so Task 1 is identical for all methods.

### 3.8 Evaluation metrics

Let At, j be the accuracy of the model after task t on E j ( j ≤ t),

$$A_{t,j} = \frac{1}{|\mathcal{E}_j|} \sum_{(x,y)\in\mathcal{E}_j} \mathbb{I}\left[\mathrm{argmax}_k\, f_{\theta_t}(x)_k = y\right] \quad (11)$$

Because E j is cumulative, At, j measures the ability to separate all classes up to task j, not only the newest one. We report:

$$AA = \frac{1}{T}\sum_{t=1}^{T} A_{t,t} \quad (12)$$

$$FA = A_{T,T} \quad (13)$$

$$FR = \frac{1}{T-1}\sum_{j=1}^{T-1} A_{T,j} \quad (14)$$

$$F = \frac{1}{T-1}\sum_{j=1}^{T-1} \left(\max_{j\le k\le T} A_{k,j} - A_{T,j}\right) \quad (15)$$

$$BWT = \frac{1}{T-1}\sum_{j=1}^{T-1} (A_{T,j} - A_{j,j}) \qquad (16)$$

We also report the macro-averaged F1 score across the 10 classes, along with the overall test accuracy is FA. of the final model. Accuracy is mostly influenced by Normal traffic, which is why macro-F1 is also reported.

# 4 Experimental Setup

## 4.1 Datasets and tasks

Table 1 summarises the datasets. In all cases the target The attack type is removed, as is the identifiers and free-text fields, numeric features are median imputed and standardised, categorical features are most-frequently-imputed and one-hot encoded, and every Only the training split is equipped with a transformer. Splits are stratified (65% train, 15% validation, 20% test, seed 42) and identical for all methods.

## 4.2 Implementation and hyper parameters

The hyper-parameters (Table 2) were kept constant during the development process. The model has been reused without any changes for the ToN-IoT and CIC-IDS2017. Experiments have been conducted using PyTorch [22] and scikit-learn. This is a single NVIDIA Tesla T4 GPU run on seed 42, as described in [23].

## 4.3 Compared conditions

**Fine-tuning (Baseline):** Following training with the No regularisation, no replay and weighted loss by the class. **EWC + replay and LwF + replay:** the methods of Section 3 trained on $D_{new\ t} \cup B_t$ . On UNSW-NB15 we also performed EWC and LwF w/o replay for the ablation in Section 5.2.

# 5 Results

## 5.1 Main comparison

Table 3 and Figure 5 report the results. Fine-tuning ends with a final accuracy between 23.6% and 36.5%

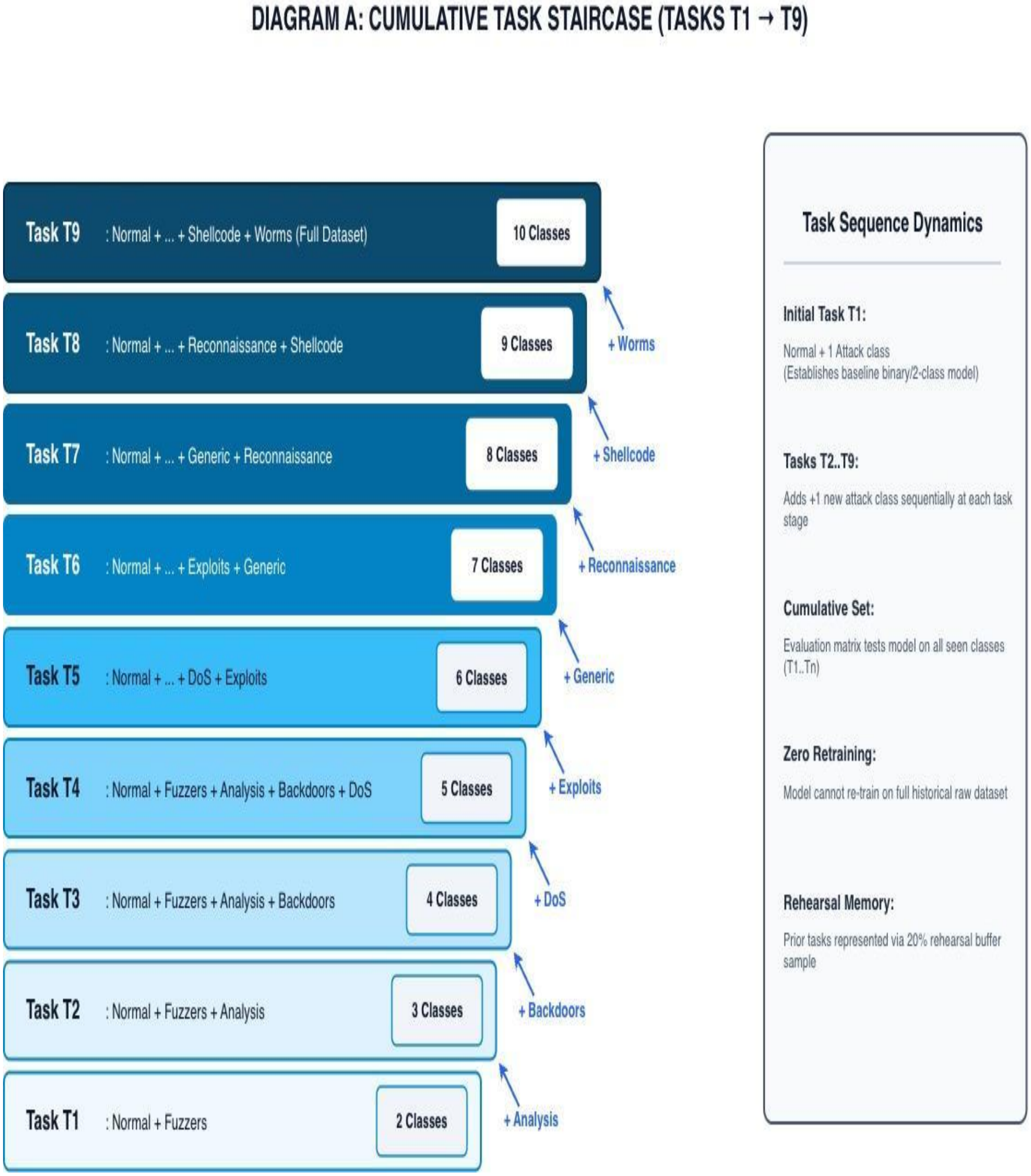

**Figure 4: Tasks (1-9)**

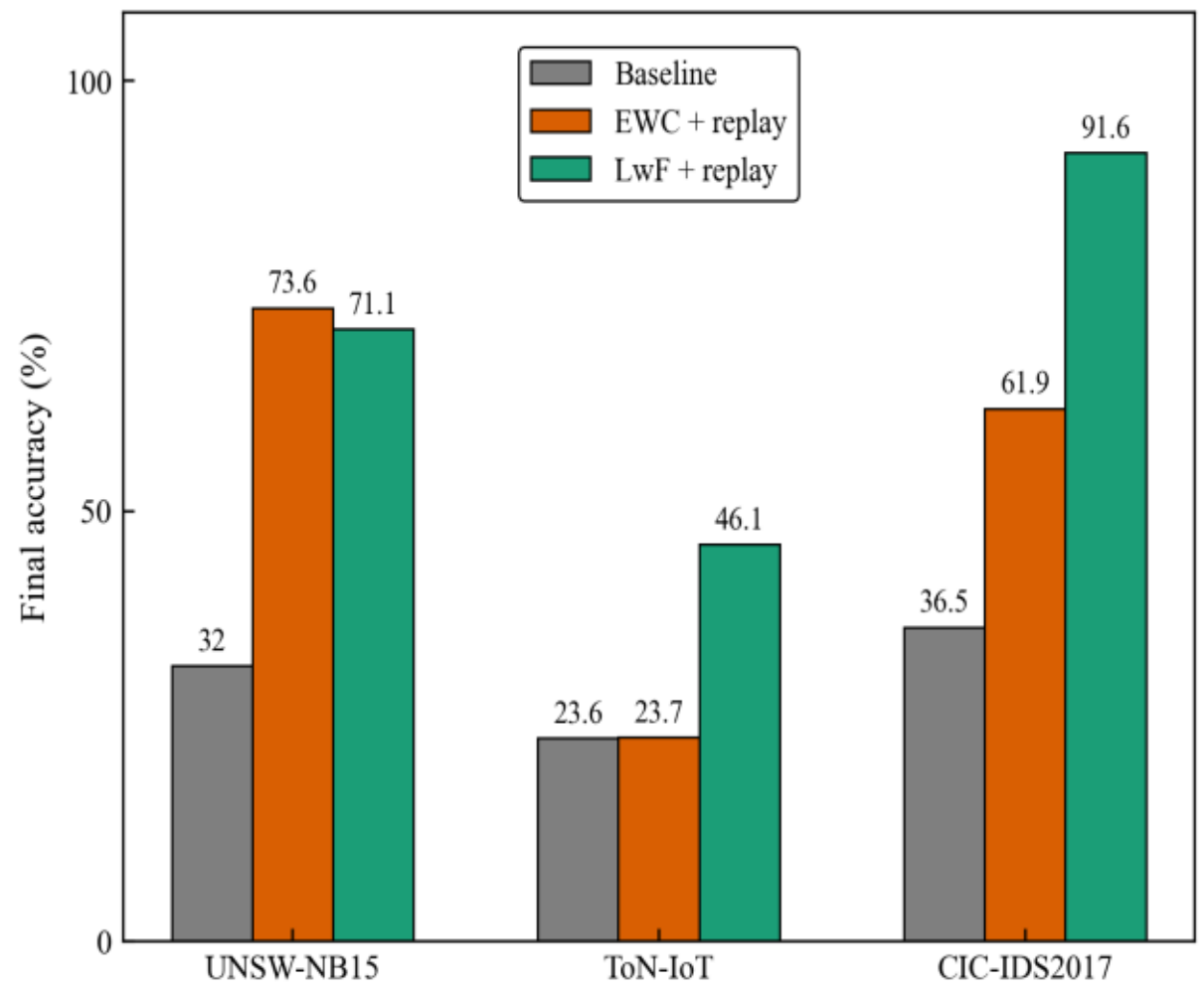

**Figure 5:** Final accuracy after the ninth task. No replay buffer is used for the Baseline, EWC and LwF use the rehearsal buffer of Eq. (6). The best method differs by dataset.

and its predictions are accurate to at least an a macro-F1 of 0.07 on all datasets: The course essentially is Normal with the latest class. On UNSW-NB15, EWC and The final accuracies of LwF with replay are 73.6% and 71.1%, respectively. retention of 72.8% and 68.8%, against 32.0% and 44.5% for fine-tuning. On ToN-IoT the final accuracy for LwF with replay is 46.1% Meanwhile, the retention rates reach 27.5% and 56.5% for EWC with replay, respectively, while they stall at 23.7% and 56.5% for EWC without replay. indistinguishable from fine tuning (23.6%). At CIC-IDS2017, LwF with replay achieved 91.6% final accuracy, 92.9% retention and macro-F1 score. of 0.77, against 61.9%, 65.2% and 0.49 for EWC and 36.5%, 45.6% and 0.07 for fine-tuning.

## 5.2 Effect of replay (UNSW-NB15 ablation)

The effect of replay from that of the other variables is separated in Table 4. the regularisers from UNSW-NB15. If not played back, EWC and LwF reach the end of: The final accuracy reached was 32.0% and 32.3% which is about the same. Fine-tuning (32.0%), with the same final retention (44.5%); LwF Increases the average accuracy (56.3% vs. 52.9%) and also increases forgetting. If there is an adaptive rehearsal buffer then final accuracy increases to 73.6% and 71.1%. N = 50 samples/ class, run for Only two epochs, provided an intermediate result (54.3% and 53.8%), which suggests a dependence on buffer size, although the different A clean comparison is not possible with epoch budget. This is a first run without, All three methods ended up at 31.9% by predicting. Close to normal for nearly all flow. We found no evidence to suggest that we should not have used the no-replay ablation.The conference will be held in conjunction with ToN-IoT or CIC-IDS2017.

## 5.3 Retention across tasks

The accuracy of the final model on each of the cumulative evaluation sets is shown in figure 6. Early sets are rapidly degraded when fine-tuned. The two regularisers have a range between approx. 0.64 and 0.81 for UNSW-NB15. LwF maintains a stable value above 0.91 on all sets; EWC is stable around 0.60-0.63 on all but the first two sets. On ToN-IoT, EWC’s last model is nearing fine tuning on each set. The step by step values behind this are informative: EWC on ToN-IoT got $A_{8,8}$ = 0.825 after 8 tasks and $A_{9,9}$ = 0.237 after 9 tasks where the least frequent class (MITM, about 730 training samples) was added.

EWC's At,t was 0.98 after Task 6 and then 0.58, 0.75 and 0.62 for the three tasks having the smallest number of new classes (Bot, Web Brute Force, Web XSS) on CIC-IDS2017; while LwF's At, t never dropped below 0.71.

### 5.4 Per-class results

The per-class final F1 of the models is shown in Table 5. For fine-tuning only those classes labeled as Normal and the last class in each of the datasets are recognized, for EWC on ToN-IoT only Normal is recognized, while there are others whose rate is almost zero in the case of best method: Backdoor and Worms on UNSW-NB15; Backdoor, Password, Scan- ning and Injection on ToN-IoT; Web Brute Force and (in precision) Web XSS on CIC-IDS2017. On CIC-IDS2017, LwF's Web XSS has recall 0.91 but precision 0.30 while Web Brute Force has recall 0.09, which is consistent with Brute Force flows being labelled as XSS; we did not verify this with the confusion matrix.

**Table 1: Datasets and task construction.** Task orders place the rarest class last.

| | UNSW-NB15 | ToN-IoT | CIC-IDS2017 |
|---|---|---|---|
| Records used | 175 341 (official training-set partition) | 211 043 (network train/test file) | 138 253 after filtering |
| Train / val / test | 113 971 / 26 301 / 35 069 | 137 177 / 31 656 / 42 210 | 89 864 / 20 738 / 27 651 |
| Input features | 194 (39 numeric, 3 categorical before one-hot) | 92 (14 numeric, 18 categorical before one-hot) | 76 (all numeric) |
| Normal flows | 56 000 | 50 000 | 50 000 (capped) |
| Rarest class | Worms (130) | MITM (1 043) | Web XSS (652) |
| Task order ($a_1, \dots, a_9$) | DoS, Exploits, Fuzzers, Reconnaissance, Generic, Analysis, Backdoor, Shellcode, Worms | Backdoor, DDoS, DoS, Injection, Password, Scanning, Ransomware, XSS, MITM | DoS Hulk, DDoS, PortScan, FTP-Patator, SSH-Patator, DoS GoldenEye, Bot, Web Brute Force, Web XSS |
| Removed columns | record id, binary label | IP addresses, ports, DNS/SSL/HTTP free-text fields, binary label | Destination Port, duplicate header-length column |
| Other handling | – | "-" placeholders set to missing | ±∞ set to missing; ≤ 20 000 flows per label per file; Monday (benign only) file excluded; Infiltration, SQL Injection, Heartbleed (too few samples to split), Slowloris and Slowhttptest dropped |

**Table 2:** Hyper-parameters (identical on all three datasets)

| Hyper-parameter | Value |
|---|---|
| Optimiser | Adam, lr 10-3 |
| EWC strength λ | 3 |
| Batch size | 32 |
| LwF weight α | 0.4 |
| LwF temperature τ | 3 |
| Max epochs | 60 |
| Early-stopping patience | 10 |
| Replay fraction γ | 0.20 |
| Random seed | 42 |
| Replay floor (Nmin) | 20 |
| Replay cap (Nmax) | 300 |

## 6 Discussion

**Replay is the decisive ingredient on UNSW-NB15 (RQ1).** The two regularisers alone were able to perform well in the single dataset we tested ablation in, and the same approach with a small replay bud (RQ1). The two regularisers alone were able to perform well in the single dataset we tested ablation in, and the same approach with a small replay budget performed a similar task that was more than 38 percentage points of final accuracy. One possible interpretation is that if a new class is the only one in the training set, and the output layer is tied to the Normal class, the shared output layer will be pulled towards the new class more quickly than the weight or output constraints can prevent, while the replay samples will maintain a direct gradient signal on the old class. We didn't test this mechanism, and we only tried two regularisation strengths without replay, so we are unable to make a statement that no strength would help.

**Table 3:** Main results (single run, seed 42; values in %, macro-F1 on a 0–1 scale). AA: mean of At,t ; FA: final accuracy; FR: final retention; F: average forgetting; BWT: backward transfer. Bold marks the best AA, FA, FR and macro-F1 per dataset. F and BWT are not bolded because they can flatter a weak model (Section 6).

| Dataset | Method | AA | FA | FR | F | BWT | macro-F1 |
|---|---|---|---|---|---|---|---|
| UNSW-NB15 | Baseline | 52.9 | 32.0 | 44.5 | 11.0 | −11.0 | 0.05 |
| | EWC + replay | 75.2 | 73.6 | 72.8 | 4.7 | −2.6 | 0.46 |
| | LwF + replay | 78.0 | 71.1 | 68.8 | 10.3 | −10.1 | 0.44 |
| ToN-IoT | Baseline | 52.7 | 23.6 | 39.6 | 16.8 | −16.8 | 0.05 |
| | EWC + replay | 76.9 | 23.7 | 40.5 | 46.3 | −43.0 | 0.04 |
| | LwF + replay | 78.6 | 46.1 | 56.5 | 31.2 | −26.2 | 0.32 |
| CIC-IDS2017 | Baseline | 54.4 | 36.5 | 45.6 | 11.0 | −11.0 | 0.07 |
| | EWC + replay | 85.8 | 61.9 | 65.2 | 27.7 | −23.6 | 0.49 |
| | LwF + replay | 91.0 | 91.6 | 92.9 | 1.6 | +2.0 | 0.77 |

**Table 4:** Effect of replay on UNSW-NB15 (values in %). The two-epoch row is shown for trend only because its epoch budgets differ.

| Replay setting | Method | AA | FA | FR | F |
|---|---|---|---|---|---|
| None (60 epochs) | Baseline | 52.9 | 32.0 | 44.5 | 11.0 |
| | EWC | 52.8 | 32.0 | 44.5 | 10.9 |
| | LwF | 56.3 | 32.3 | 44.5 | 14.8 |
| Fixed, 50 per class (2 epochs) | Baseline | 52.3 | 31.9 | 44.5 | 10.3 |
| | EWC | 62.1 | 54.3 | 56.6 | 6.7 |
| | LwF | 64.7 | 53.8 | 55.6 | 10.4 |
| Adaptive, Eq. (6) (60 epochs) | Baseline | 52.9 | 32.0 | 44.5 | 11.0 |
| | EWC | 75.2 | 73.6 | 72.8 | 4.7 |
| | LwF | 78.0 | 71.1 | 68.8 | 10.3 |

**Which regulariser wins depends on the dataset (RQ2).** EWC was slightly ahead on UNSW-NB15 (73.6% vs. 71.1%) a difference that cannot be differentiated from noise), but LwF was significantly outdone (with only a single seed) on ToN-IoT and CIC-IDS2017. The failures of EWC are concentrated in the last tasks, where the new class is small (Section 5). This may be because the Fisher information is not normalised across tasks, meaning that the effective penalty increases as the number of tasks increases and decreases the plasticity required to learn a small new class, while LwF only constrains the output of the old classes, leaving the weights more free. This is a hypothesis but we have not changed λ per data set, nor have we normalised the Fisher estimates. We also periodically noticed that ToN-IoT and CIC-IDS2017 exhibited a greater degree of validation loss, which is something much less frequent on UNSW-NB15 - it could be a result of the penalty and both gradient clipping and decreasing the learning rates are obvious follow-ups.

**Some metrics mislead (RQ3).** The forgetting of the previous classes already happened in task j and is contained in A j, j, since the evaluation sets E j are cumulative, and the model is single-head. Therefore, for further fine tuning, A j, j is comparatively small from the third task on (0.33 on UNSW-NB15 at Task 6) and a small increase thereof would still allow significant improvement in the forgetting measure value: A j, j for the baseline (0.33 on UNSW-NB15 at Task 6) is lower than EWC's value on the considered tasks (ToN-IoT and CIC-IDS2017) despite its much poorer final accuracy. Most of EWC on ToN-IoT (46.3%) is forgotten due to high At,t until the last task. The ambiguous case of backward transfer is also observed: The positive BWT score (+2.0) of the weak intermediate model CIC-IDS2017 (A8,8 =

0.717) is recovered after the final task, AT,8 = 0.916; this is perhaps more plausibly interpreted as recovery from an under-trained model than as genuine backward transfer. Thus, we suggest using forgetting and BWT together when assessing final accuracy, retention, and macro-F1 and caution the use of forgetting alone for ranking.

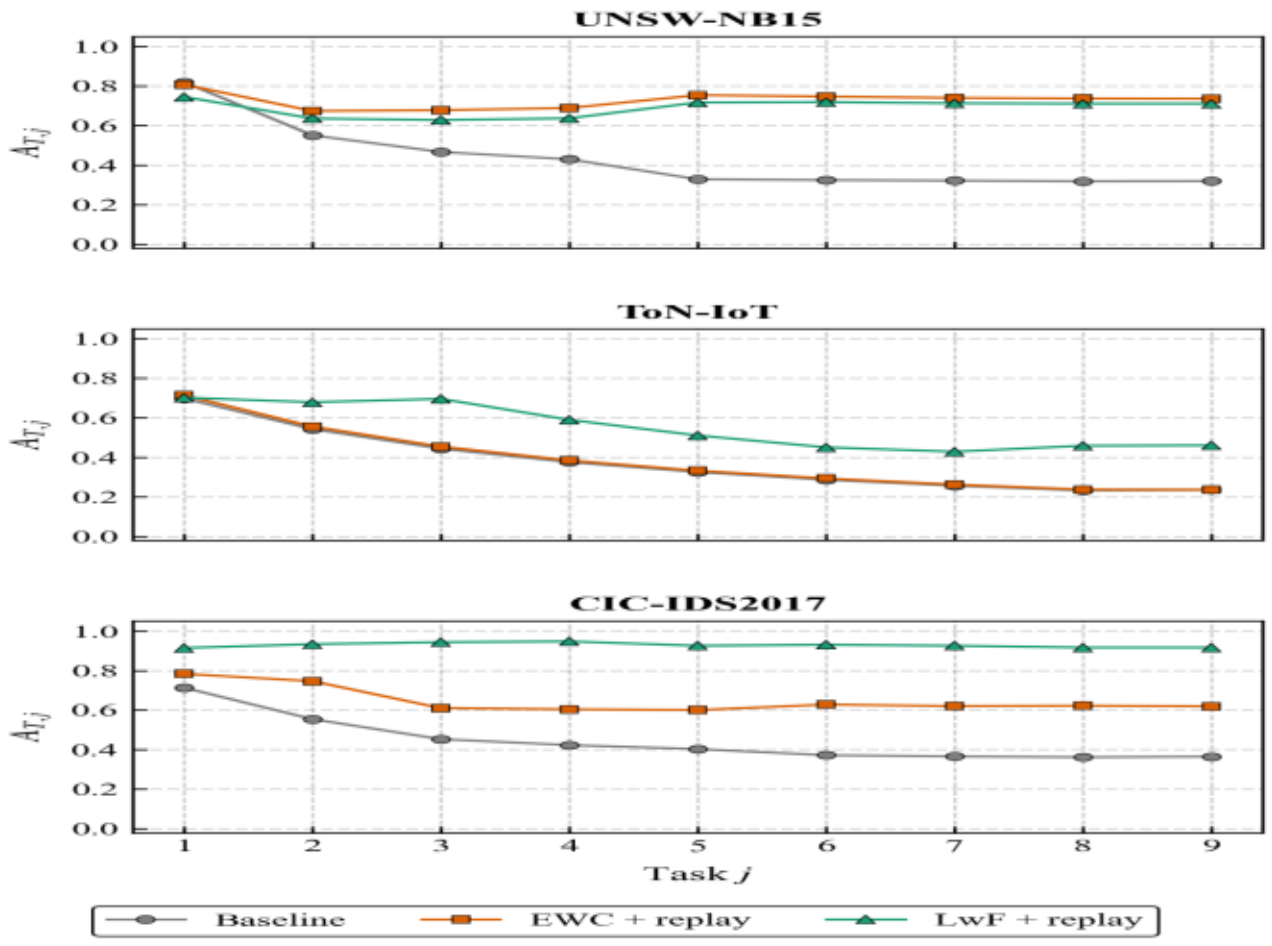

**Figure 6:** Retention profile.Accuracy AT, j of the final model on the cumulative test set of each task j. Higher and flatter is better

**Table 5:** Per-class F1 of the final models (B: baseline, E: EWC + replay, L: LwF + replay). Classes are listed in task order.

| UNSW-NB15 | | | | ToN-IoT | | | | CIC-IDS2017 | | | |
|---|---|---|---|---|---|---|---|---|---|---|---|
| Class | B | E | L | Class | B | E | L | Class | B | E | L |
| DoS | .00 | .44 | .10 | Backdoor | .00 | .00 | .00 | DoS Hulk | .00 | .47 | .87 |
| Exploits | .00 | .54 | .56 | DDoS | .00 | .00 | .72 | DDoS | .00 | .76 | .98 |
| Fuzzers | .00 | .60 | .59 | DoS | .00 | .00 | .48 | PortScan | .00 | .00 | .99 |
| Reconnaissance | .00 | .60 | .72 | Injection | .00 | .00 | .01 | FTP-Patator | .00 | .67 | 1.00 |
| Generic | .00 | .98 | .99 | Password | .00 | .00 | .00 | SSH-Patator | .00 | .63 | .66 |
| Analysis | .00 | .24 | .12 | Scanning | .00 | .00 | .00 | DoS GoldenEye | .00 | .94 | .97 |
| Backdoor | .00 | .00 | .08 | Ransomware | .00 | .00 | .39 | Bot | .00 | .03 | .73 |
| Shellcode | .00 | .27 | .40 | XSS | .00 | .00 | .80 | Web Brute Force | .00 | .73 | .15 |
| Worms | .03 | .00 | .00 | MITM | .03 | .00 | .24 | Web XSS | .14 | .07 | .45 |
| Normal | .50 | .88 | .90 | Normal | .49 | .38 | .56 | Normal | .55 | .66 | .91 |
| Macro-F1 | .05 | .46 | .44 | Macro-F1 | .05 | .04 | .32 | Macro-F1 | .07 | .49 | .77 |

Accuracy hides rare-class failure. Macro-F1 and per-class scores are crucial due to the overall accuracy that is dominated by Normal traffic. Even the most optimal arrangements fail to detect a few classes of attacks (Table 5). For these types of intrusion detectors, it's the expensive mistake, as a missed attack class is worse than a handful of extra Alarms.

## 7 Limitations

**1. No replay-only baseline.** EWC and LwF received a replay while the fine-tuning baseline did not receive any replay. The gains on ToN-IoT and CIC-IDS2017 for the regularisers are then just due to replay, and it is uncertain if EWC or LwF increases the gains beyond replay. The UNSW-NB15 ablation reveals that regularisers w/o replay do not help there, not that they help beyond replay.

**2. Single seed.** All numbers are from one run, using seed 42. No variance was reported and small differences and small gaps (for example 73.6% vs. 71.1% on UNSW-NB15) should not be taken as significant.

**3. Configuration selected on UNSW-NB15 test results.** The numbers for UNSW-NB15 are based on tests split and results may be optimistic because a separate model selection split wasn't done. ToN-IoT and CIC-IDS2017 were executed in the frozen configuration which is the fairer out-of-sample check, however the parameters $\lambda$ , $\alpha$ and the buffer size were not tuned for these two checks.

**4. Replay implementation.** At each task, the buffer is re-drawn from the previously stored training samples, which allows the buffer to grow beyond the length of an arbitrary memory. Normal traffic is also enabled at all tasks.

**5. Early stopping ignores old classes.** Check-point selection is not based on the retention rate, but rather on the validation samples of new class and Normal only.

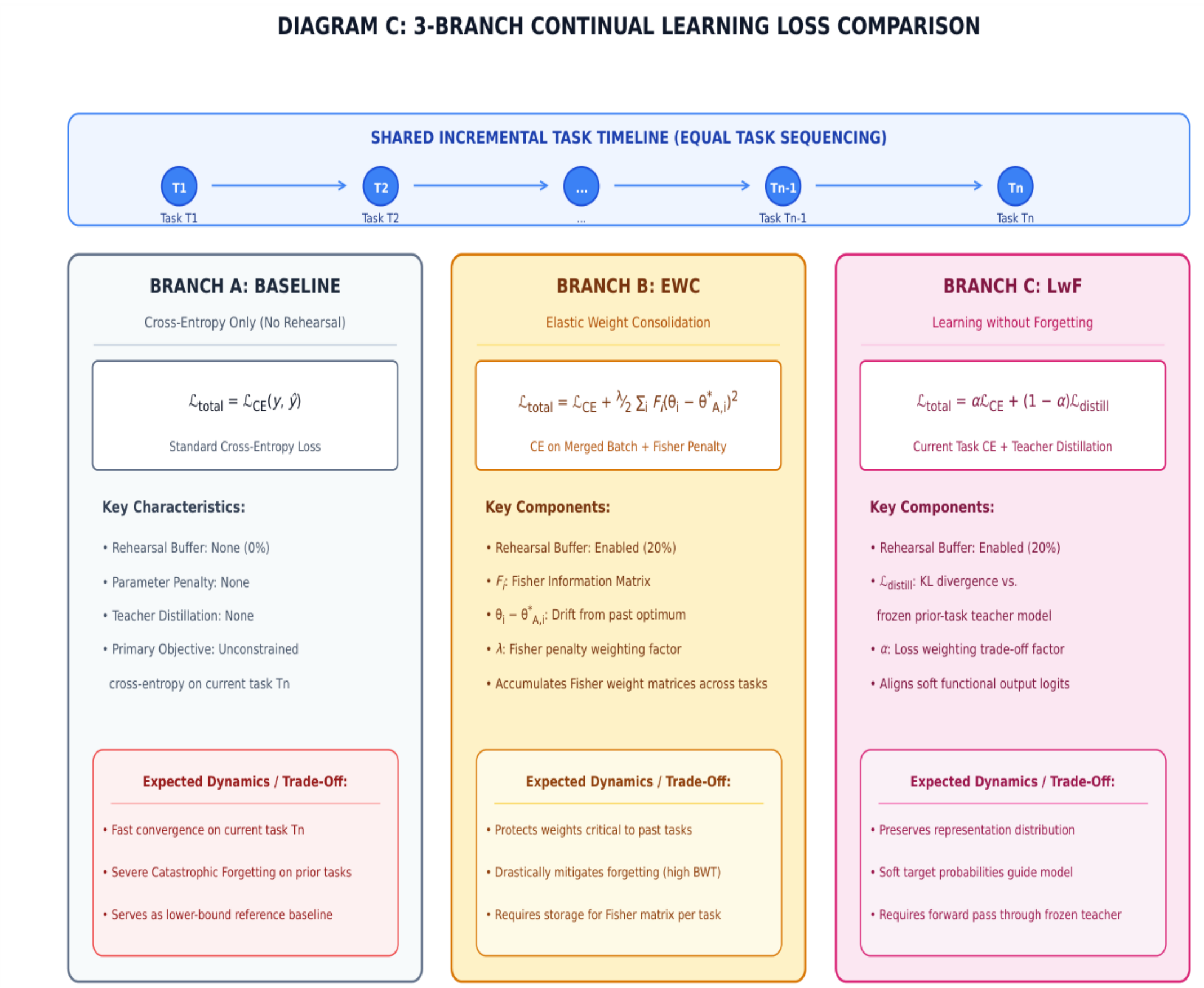

**Figure 7: Loss comparison**

**6. One task order per dataset**. The rarest class was the last class. Order effects were not examined and the relatively poor performance on the last-rare tasks could partly be due to this.

**7. Dataset preparation.** CIC-IDS2017 was sub-sampled (per file label cap, Normal cap) and some classes were dropped. The class counts used do not reflect the natural traffic mix of the data set used in these two cases, as documented by ToN-IoT and CIC-IDS2017 [17]. Results are offline and flow-based, with no concept drift, adversarial inputs or latency constraints.

## 8 Conclusion and Future Work

The small rehearsal buffer, which adapts to class size, changed EWC and LwF from methods that were indistinguishable from fine-tuning into methods that kept most of the earlier classes on UNSW-NB15 (final accuracy 73.6%, 71.1% compared to 32.0%). LwF with replay was the best on ToN-IoT (46.1%) and CIC-IDS2017 (91.6%), while EWC with replay was competitive on UNSW-NB15 and did not perform well on the remaining, small tasks of the other two datasets (23.6% and 36.5%). The strength of the regularisers was then dependent on the dataset, and the strength optimized on one dataset couldn't be transferred to other ones. The forgetting and backward-transfer measures should be interpreted in conjunction with the final accuracy, retention and macro-F1 and are susceptible to weak models. First, a baseline should be added, the replay-only, which is by far the most important missing control, and all experiments should be repeated on several seeds with confidence intervals, in the future. Other steps include: a held out validation protocol where old classes are used to choose hyper-parameters and to stop training early; sweeps across λ , α and buffer size on each dataset; a fixed-size memory that is kept to a reasonable size with principled selection rule; normalised or decayed Fisher estimates; replay and EWC combined; other orders of tasks; and attack classes that are not labelled and therefore would not have been replayed.

## CRediT authorship contribution statement

[Mohammed Fackruddin Tusher]: Conceptualization, Methodology, Software, Investigation, Formal analysis, Visualization, Writing – original draft.
[Muhammad Iqbal Hossain]: Supervision, review.

## Declaration of competing interest

The authors do not have any supporting financial or other interests that would appear to have influenced the work or interpretation of this paper.

## Data and code availability

UNSW-NB15, ToN-IoT and CIC-IDS2017 are publicly available from their original providers [14–16]. The experiment code and the saved result tables (accuracy matrices, per-task retention, final classification reports) are available at [repository URL].